\pdfoutput=1
\documentclass[conference,a4paper,9pt]{IEEEtran}
\usepackage{amsmath,amssymb,amsfonts}
\usepackage{mathtools}
\usepackage{bm}
\usepackage{algorithmic}
\usepackage{graphicx}
\usepackage{textcomp}
\usepackage{xcolor}
\usepackage[acronym,toc]{glossaries}
\usepackage{cleveref}
\usepackage{siunitx}
\usepackage{tabularx}
\usepackage{booktabs}
\usepackage{pgfplots, tikz}
\pgfplotsset{compat=1.17}
\usetikzlibrary{arrows.meta}
\usepackage{tikzit}
\usepackage{layouts}
\usepackage{environ}
\usepackage{placeins}
\ifCLASSOPTIONcompsoc
    \usepackage[caption=false, font=normalsize, labelfont=sf, textfont=sf]{subfig}
\else
    \usepackage[caption=false, font=footnotesize]{subfig}
\fi
\usepackage[activate={true,nocompatibility},
    final,
    tracking=true,
    kerning=true,
    spacing=true,
    factor=1100,
    stretch=25,
    shrink=25,
    letterspace=-30
]{microtype}

\makeatletter
\newsavebox{\measure@tikzpicture}
\NewEnviron{scaletikzpicturetowidth}[1]{%
    \def\tikz@width{#1}%
    \begin{lrbox}{\measure@tikzpicture}%
        \BODY
    \end{lrbox}%
    \pgfmathparse{#1/\wd\measure@tikzpicture}%
    \BODY
}
\makeatother

\newacronym{admm}{ADMM}{Alternating Directions of Multipliers Method}
\newacronym{anm}{ANM}{Atomic Norm Minimization}
\newacronym{adc}{ADC}{Analog-to-Digital Converter}
\newacronym{awgn}{AWGN}{Additive White Gaussian Noise}
\newacronym{asic}{ASIC}{Application Specific Integrated Circuit}
\newacronym{arpack}{ARPACK}{ARnoldi PACKage}
\newacronym{api}{API}{Application Programmable Interface}
\newacronym{aut}{AUT}{Antenna Under Test}
\newacronym[plural=AOI, firstplural=Areas of Interest (AOI)]{aoi}{AOI}{Area of Interest}
\newacronym{ai}{AI}{Artifical Intelligence}
\newacronym{aic}{AIC}{Akaike Information Criterion}

\newacronym{bp}{BP}{Basis Pursuit}
\newacronym{bpdn}{BPDN}{Basis Pursuit Denoising}
\newacronym{blue}{BLUE}{Best Linear Unbiased Estimator}

\newacronym{cnn}{CNN}{Convolutional Neural Network}
\newacronym{crb}{CRB}{Cramér-Rao Bound}
\newacronym{cs}{CS}{Compressed Sensing}
\newacronym{cf}{CF}{Crest factor}
\newacronym{cr}{CR}{Compression Ratio}
\newacronym{cmos}{CMOS}{Complementary Metal Oxide Semiconductor}
\newacronym{cots}{COTS}{Commercial-Off-The-Shelf}
\newacronym{cpu}{CPU}{Central Processing Unit}
\newacronym{cfar}{CFAR}{Constant False Alarm Rate}
\newacronym{cvx}{CVX}{Convex Optimization toolboX}

\newacronym{doa}{DoA}{Direction of Arrival}
\newacronym{dod}{DoD}{Direction of Departure}
\newacronym[plural=DMC,firstplural=Dense Multipath Components (DMC)]{dmc}{DMC}{Dense Multipath Components}
\newacronym{das}{DAS}{Delay-and-Sum}
\newacronym{dft}{DFT}{Discrete Fourier Transform}
\newacronym{idft}{IDFT}{Inverse Discrete Fourier Transform}
\newacronym{dct}{DCT}{Discrete Cosine Transform}
\newacronym{dsp}{DSP}{Digital Signal Processor}
\newacronym{dnn}{DNN}{Deep Neural Network}

\newacronym{eadf}{EADF}{Effective Aperture Distribution Function}
\newacronym{etadf}{ETADF}{Effective Time-Aperture Distribution Function}
\newacronym{esprit}{ESPRIT}{Estimation of Signal Parameters via Rotational Invariance Techniques}
\newacronym{ett}{ETT}{Eigenvalue Threshold Test}
\newacronym{edc}{EDC}{Efficient Detection Criterion}
\newacronym{eft}{EFT}{Exponential Fitting Test}

\newacronym{fc}{FC}{fully-connected}
\newacronym{fht}{FHT}{Fast Hadamard Transform}
\newacronym[longplural={Fast Fourier Transforms}]{fft}{FFT}{Fast Fourier Transform}
\newacronym{fmcw}{FMCW}{Frequency-Modulated Continuous-Wave}
\newacronym{fpga}{FPGA}{Field Programmable Gate Array}
\newacronym{fri}{FRI}{Finite Rate of Innovation}
\newacronym{fim}{FIM}{Fisher Information Matrix}
\newacronym{fmc}{FMC}{Full Matrix Capture}
\newacronym{fista}{FISTA}{Fast Iterative Shrinkage-Thresholding Algorithm}
\newacronym{frvm}{FRVM}{Fast Relevance Vector Machine}

\newacronym{gfcs}{grid-free CS}{grid-free compressive sensing}
\newacronym{gpu}{GPU}{Graphical Processing Unit}
\newacronym{gtd}{GTD}{Geometrical Theory of Diffraction}
\newacronym{gan}{GAN}{Generative Adversarial Network}

\newacronym{hrpe}{HRPE}{High Resolution Parameter Estimator}
\newacronym{hfss}{Ansys HFSS}{Ansys High Frequency Electromagnetic Simulation Software}

\newacronym[longplural={Inverse Fast Fourier Transforms}]{ifft}{IFFT}{Inverse Fast Fourier Transform}
\newacronym{ir}{IR}{Impulse Response}
\newacronym{iid}{iid}{independent and identically distributed}
\newacronym{iir}{IIR}{Infinite Impulse Response}
\newacronym{irf}{IRF}{Impulse Response Function}
\newacronym{ista}{ISTA}{Iterative Shrinkage-Thresholding Algorithm}

\newacronym{kld}{KLD}{Kullback-Leibler Divergence}

\newacronym{ls}{LS}{Least Squares}
\newacronym{lasso}{LASSO}{Least Absolute Shrinkage and Selection Operator}
\newacronym{lse}{LSE}{Line Spectral Estimation}
\newacronym{lfsr}{LFSR}{Linear Feedback Shift Register}
\newacronym{lo}{LO}{Local Oscillator}
\newacronym{los}{LOS}{Line of Sight}
\newacronym{lti}{LTI}{Linear Time-Invariant}
\newacronym{lam}{LAM}{Large Area Monitoring}

\newacronym{mimo}{MIMO}{multiple input multiple output}
\newacronym{mmv}{MMV}{multiple measurement vectors}
\newacronym{ml}{ML}{maximum likelihood}
\newacronym{mmse}{MMSE}{misspecified mean squared error}
\newacronym{mse}{MSE}{mean squared error}
\newacronym{bce}{BCE}{Binary Crossentropy}
\newacronym{mlbs}{MLBS}{Maximum Length Binary Sequence}
\newacronym{mpc}{MPC}{Multipath Component}
\newacronym{msm}{MSM}{M-Sequence Method}
\newacronym{mwc}{MWC}{Modulated Wideband Converter}
\newacronym{mpm}{MPM}{Matrix Pencil Method}
\newacronym{mpu}{MPU}{Microprocessor Unit}
\newacronym{mri}{MRI}{Magnetic resonance imaging}
\newacronym{music}{MUSIC}{Multiple Signal Classification}
\newacronym{mkl}{MKL}{Math Kernel Library}
\newacronym{mcrb}{MCRB}{Misspecified Cramér-Rao Bound}
\newacronym{mmle}{MMLE}{Misspecified Maximum-Likelihood Estimator}
\newacronym{ndt}{NDT}{Nondestructive Testing}
\newacronym{nde}{NDE}{Nondestructive Evaluation}
\newacronym{nn}{NN}{Neural Net}
\newacronym{nist}{NIST}{National Institute of Standards and Technology}

\newacronym{omp}{OMP}{Orthogonal Matching Pursuit}
\newacronym{oop}{OOP}{Object Oriented Programming}

\newacronym{pdp}{PDP}{Power Delay Profile}
\newacronym{pn}{PN}{Pseudo-Noise}
\newacronym{pwc}{PWC}{Plane Wave Compounding}
\newacronym{pwi}{PWI}{Plane Wave Imaging}
\newacronym{pura}{PURA}{Patch Uniform Rectangular Array}
\newacronym{pymax}{PyMAX}{Python Maximization Approach}
\newacronym{pts}{PTS}{Pseudo-True Solution}
\newacronym{pdf}{pdf}{probability density function}

\newacronym{relu}{ReLU}{Rectified Linear Unit}
\newacronym{resnet}{ResNet}{Residual Neural Network}
\newacronym{ram}{RAM}{Random Access Memory}
\newacronym{rcs}{RCS}{Radar Cross Section}
\newacronym{rd}{RD}{Random Demodulator}
\newacronym{rx}{RX}{receiver}
\newacronym{rem}{REM}{Reconstruction Error Metric}
\newacronym{rmse}{RMSE}{Root Mean Squared Error}
\newacronym{rms}{RMS}{root mean squared}
\newacronym{ric}{RIC}{Restricted Isometry Constant}
\newacronym{rip}{RIP}{Restricted Isometry Property}
\newacronym{rc}{RC}{Raised Cosine}
\newacronym{roi}{ROI}{Region of Interest}
\newacronym{rimax}{RIMAX}{Richter Maximization Approach}
\glsunset{rimax}
\newacronym{rvm}{RVM}{Relevance Vector Machine}

\newacronym{scf}{SCF}{spatial correlation function}
\newacronym{samurai}{SAMURAI}{Synthetic Aperture Measurements of Uncertainty in Angle of Incidence}
\newacronym[plural=SC,firstplural=Specular Components (SC)]{sc}{SC}{Specular Components}
\newacronym{sdp}{SDP}{semi-definite program}
\newacronym{sdr}{SDR}{Signal to Diffuse Ratio}
\newacronym{simd}{SIMD}{Single Instruction Multiple Data}
\newacronym{svd}{SVD}{singular value decomposition}
\newacronym{svm}{SVM}{Support Vector Machine}
\newacronym{soe}{SOE}{Sparsity Order Estimation}
\newacronym{sgd}{SGD}{Stochastic Gradient Descent}
\newacronym{stuca}{StUCA}{Stacked Uniform Circular Array}
\newacronym{spucpa}{SPUCPA}{Stacked Polarimetric Uniform Circular Patch Array}
\newacronym{suca}{SUCA}{Stacked Uniform Circular Array}
\newacronym{saft}{SAFT}{Synthetic Aperture Focusing Technique}
\newacronym{sota}{SOTA}{State of the Art}
\newacronym{ssd}{SSD}{Solid State Device}
\newacronym{ssr}{SSR}{Sparse Signal Recovery}
\newacronym{sa}{SA}{Synthetic Aperture}
\newacronym{spw}{SPW}{Single Plane Wave}
\newacronym{shm}{SHM}{Structural Health Monitoring}
\newacronym{snr}{SNR}{Signal-to-Noise Ratio}
\newacronym{stela}{STELA}{Soft-Thresholding with Exact Line Search Algorithm}
\newacronym{siso}{SISO}{Single Input Single Output}
\newacronym{simo}{SIMO}{Single Input Multiple Output}
\newacronym{swe}{SWE}{Spherical Wave Expansion}
\newacronym{sme}{SME}{Spherical Mode Expansion}
\newacronym{sage}{SAGE}{Space-Alternating Generalized Expectation-Maximization}

\newacronym{th}{T\&H}{Track and Hold}
\newacronym{tf}{TF}{Transfer Function}
\newacronym{tx}{TX}{transmitter}
\newacronym{twista}{TWISTA}{Two-step Iterative Shrinkage-Thresholding Algorithm}
\newacronym{tof}{ToF}{Time-of-Flight}

\newacronym{uca}{UCA}{uniform circular array}
\newacronym{ura}{URA}{Uniform Rectangular Array}
\newacronym{ula}{ULA}{Uniform Linear Array}
\newacronym{uwb}{UWB}{Ultra-Wideband}
\newacronym{usndt}{US-NDT}{Ultrasonic Non-destructive Testing}

\newacronym{vna}{VNA}{Vector Network Analyser}
\newacronym{vsh}{VSH}{Vector Spherical Harmonics}
\def\BibTeX{{\rm B\kern-.05em{\sc i\kern-.025em b}\kern-.08em
T\kern-.1667em\lower.7ex\hbox{E}\kern-.125emX}}

\newcommand{\N}{\mathbb{N}}
\newcommand{\R}{\mathbb{R}}
\newcommand{\C}{\mathbb{C}}

\newcommand{\Norm}[1]{{\left\Vert #1\right\Vert}}

\newcommand{\reddot}{\tikz{\fill[red, draw=black] (0,0) circle (2pt);}}
\newcommand{\whitecircle}{\tikz{\fill[white, draw=black] (0,0) circle (3pt);\fill[white, draw=black] (0,0) circle (1.6pt);}}
\newcommand{\blackcircle}{\tikz{\fill[black, draw=black] (0,0) circle (3pt);\fill[white, draw=black] (0,0) circle (2.4pt);}}

\definecolor{coolor1}{HTML}{ff9900}
\definecolor{coolor2}{HTML}{29789b}
\definecolor{coolor3}{HTML}{f1f7ed}
\definecolor{coolor4}{HTML}{1a936f}
\definecolor{coolor5}{HTML}{cf3155}

\ifx\qty\undefined
    \let\oldnum\num
    \renewcommand{\num}[2][ignored]{\oldnum{#2}}

    \ifx\qty\undefined
        \newcommand{\qty}[3][ignored]{\SI{#2}{#3}}
    \fi

    \ifx\qtyrange\undefined
        \newcommand{\qtyrange}[4][ignored]{\SIrange{#2}{#3}{#4}}
    \fi
\fi

\usepackage[
    sorting=none,
    maxbibnames=99,
    defernumbers=true,
    style=ieee,
    giveninits=true,
]{biblatex}
\AtEveryBibitem{
    \clearlist{address}
    \clearfield{url}
    \clearfield{month}
    \clearfield{editor}
    \clearfield{eprint}
    \clearfield{volume}
    \clearfield{isbn}
    \clearfield{issn}
    \clearfield{pages}
    \clearfield{addendum}
    \clearlist{location}
    \clearlist{pages}
    \clearlist{editor}
}
\makeatletter
\let\blx@rerun@biber\relax
\makeatother

\begin{document}
\title{Grid-free Harmonic Retrieval and Model Order Selection using Convolutional Neural Networks}

\author{\IEEEauthorblockN{
S. Schieler\IEEEauthorrefmark{1}, 
S. Semper\IEEEauthorrefmark{1}, 
R. Faramarzahangari\IEEEauthorrefmark{1}, 
M. D\"obereiner\IEEEauthorrefmark{2}, 
C. Schneider\IEEEauthorrefmark{1}\IEEEauthorrefmark{2},
R. Thom\"a\IEEEauthorrefmark{1}
}                                     
\IEEEauthorblockA{\IEEEauthorrefmark{1}
Technische Universit\"at Ilmenau: FG EMS, Ilmenau, Germany, steffen.schieler@tu-ilmenau.de}
\IEEEauthorblockA{\IEEEauthorrefmark{2}
Fraunhofer Institute of Integrated Circuits: Dep. EMS, Ilmenau, Germany}  
}

\maketitle

\begin{abstract}
    Harmonic retrieval techniques are the foundation of radio channel sounding, estimation and modeling. 
    This paper introduces a Deep Learning approach for joint delay- and Doppler estimation from frequency and time samples of a radio channel transfer function.
      
    Our work estimates the two-dimensional parameters from a signal containing an unknown number of paths. 
    Compared to existing deep learning-based methods, the signal parameters are not estimated via classification but in a quasi-grid-free manner. 
    This alleviates the bias, spectral leakage, and ghost targets that grid-based approaches produce. 
    The proposed architecture also reliably estimates the number of paths in the measurement. 
    Hence, it jointly solves the model order selection and parameter estimation task. 
    Additionally, we propose a multi-channel windowing of the data to increase the estimator's robustness.
    
    We also compare the performance to other harmonic retrieval methods and integrate it into an existing maximum likelihood estimator for efficient initialization of a gradient-based iteration.
\end{abstract}

\vskip0.5\baselineskip

\section{Introduction}\label{sec:introduction}
Harmonic Retrieval is a problem encountered in many signal processing tasks, e.g., channel estimation~\cite{thomae2005multidim_cs}, radar localization, and direction finding.
Available solutions for the task can be divided into four groups: subspace algorithms, like \gls{music}~\cite{schmidt1986MUSIC} or \gls{esprit}~\cite{roy1989esprit}, iterative \gls{ml}~\cite{richter_estimation_2005}, and \gls{ssr}~\cite{malioutov2005ssr_source_loc} and finally most recently \gls{dnn}-based algorithms~\cite{izacard_data-driven_2019, papageorgiou_deep_2020, chen_robust_2022, naseri_machine_2022, liu_super_2020, barthelme_machine_2021}.
Our work belongs to the latter category and presents a new approach to solve a Harmonic Retrieval task by using a deep \gls{cnn}.
We note, the applications in the listed works, namely spectral- and \gls{doa}-estimation, share the algebraic structure of Harmonic Retrieval.\par
In \cite{izacard_data-driven_2019} a \gls{cnn} is used to estimate frequencies components by predicting a super-resolution pseudo-spectrum from a superposition of up to ten components.
A separate, subsequent network performs the model order estimation task.
The results show a performance improvement when compared to \gls{music}, especially in the low-\gls{snr} domain.
In \cite{papageorgiou_deep_2020} a \gls{cnn} is trained to perform \gls{doa} estimation of up to three unknown sources by determining their location on a grid, i.e., solving a classification problem.
Similarly, the authors of \cite{chen_robust_2022} address the problem of \gls{doa} estimation by combining a denoising autoencoder with another \gls{dnn} for the estimation.
As in \cite{izacard_data-driven_2019}, both approaches show performance improvements in the low-\gls{snr} domain compared to \gls{music}.
However, classification methods suffer from an inherent estimation bias due to grid-mismatch and poor scaling in terms of desired resolution.
Since increasing the grid density increases classification complexity non-linearly.
Naturally, estimates that do not rely on a grid, i.e., classification, are highly interesting, especially if we wish to achieve super-resolution.
Comparably, the work in \cite{barthelme_machine_2021} combines a regression task solved by a \gls{dnn} with gradient steps on the likelihood function of the data to perform \gls{doa} estimation.
The grid-free estimates from the \gls{dnn} are used as initial guesses for a second-order Newton method.
This approach combines the best of two worlds: fast, robust, but approximate initial estimates combined with an iterative high-resolution method allowing quadratic convergence, but only after successful initialization provided by the \gls{dnn}.
The results highlight the decrease in computational complexity and improvements in performance from the use maximum-likelihood methods.
However, the presented method only considers up to $3$ stochastic sources estimated from multiple snapshots, which is insufficient in many practical scenarios, especially if parameters are modeled as deterministic.
More significantly, the used antenna array geometries have a very small aperture, rendering the initialization of the Newton method very well conditioned since it tolerates high deviations of the initialization from the true solution.\par
Compared to previous works, our proposed architecture can estimate up to $20$ deterministic paths from a single snapshot in terms of their delay and Doppler-shift from frequency-time data.
The number of paths is estimated together with their propagation parameters.
We exploit the sparsity of the parameter space by dividing it up into a low number of auxiliary grid-cells.
Then we employ a \gls{cnn} to estimate the number of paths in each cell and their respective deviations from the cells' centers to obtain grid-free estimates.
In effect, we solve multiple joint regression and classification tasks to obtain the number of sources and their parameters.
To render the training more robust, we apply a set of windowing functions both in frequency and time domain to the input data, effectively presenting the same data with different pulse-shapes.
To verify the performance, we consider the \gls{mse} of the raw \gls{dnn} estimates and compare them to existing methods, i.e., \gls{dft} and \gls{rimax}~\cite{richter_estimation_2005}, and the \gls{crb} as the theoretical lower bound.
We also conduct an isolated analysis of the model order selection performance and compare our proposal to \gls{rimax} and \gls{edc}.
Motivated by the approach in \cite{barthelme_machine_2021}, we also use the estimates directly to warm-start a second-order gradient iteration and show that the initial guesses allow convergence with high probability.
The results indicate that the proposed methodology can reliably predict both model order and parameters sufficiently well and with comparably short computation time.
When enhanced with a few steps of an iterative \gls{ml} estimator, the performance can be improved substantially, resulting in a well-performing, robust $2$D parameter estimator with moderate computational complexity.
\begin{figure*}[t]
    \centering
    \begin{scaletikzpicturetowidth}{\textwidth}

\tikzstyle{box}=[fill=none, draw=none, shape=rectangle, minimum width=7.5em, minimum height=3.0em, anchor=north]
\tikzstyle{container box}=[shape=rectangle, minimum width=9.6em, minimum height=22.5em, fill opacity=0.2, anchor=north]
\tikzstyle{innercontainer}=[fill=none, draw=black, shape=rectangle, minimum height=14em, minimum width=8.4em, densely dashdotted, anchor=north]
\tikzstyle{threelines}=[shape=rectangle, fill=coolor1, minimum height=4.0em, minimum width=7.5em, draw=black, fill opacity=0.69, anchor=north]
\tikzstyle{twolines}=[fill=white, draw=black, shape=rectangle, minimum width=7.5em, minimum height=2.5em, fill opacity=0.8, anchor=north]
\tikzstyle{oneline}=[fill=white, draw=black, shape=rectangle, minimum height=1.0em, minimum width=7.5em, fill opacity=0.8, anchor=north]

\tikzstyle{arrow}=[->, >=Triangle]
\tikzstyle{thickarrow}=[->, >=Triangle, thick]
\tikzstyle{thick line}=[-, thick]
\tikzstyle{thin line}=[-, thin]

\tikzstyle{arrow}=[->, >=Triangle]
\tikzstyle{thickarrow}=[->, >=Triangle, thick]
\tikzstyle{thick line}=[-, thick]

\tikzstyle{arrow}=[->, >=Triangle]
\tikzstyle{thickarrow}=[->, >=Triangle, thick]

\newcommand{\titleface}[1]{\normalsize{\textbf{#1}}}

\begin{tikzpicture}
	\begin{pgfonlayer}{nodelayer}
		\node [style=container box] (32) at (0,0)[xshift=12em, yshift=0em] {};
		\node [style=box] (24) at (0,0)[xshift=12em, yshift=-1em, label={[yshift=-1em, align=center]\titleface{Channel Upscaling}}] {};
		\node [style=threelines, label={[align=center, anchor=north]Conv2D\\\small{k{:} 3x3, s{:} 1x1,}\\\small{$C_{\text{out}}${:} $2C_{\text{in}}$}}] (29) at (0,0)[xshift=12em, yshift=-4.5em] {};
		\node [style=oneline, fill=coolor4, fill opacity=1] (30) at (0,0)[xshift=12em, yshift=-10.0em] {Batch Norm.};
		\node [style=oneline, fill=coolor2, fill opacity=1] (31) at (0,0)[xshift=12em, yshift=-13.0em] {ReLU};
		\node [style=innercontainer, minimum height=11em] (77) at (0,0)[xshift=12em, yshift=-4.0em] {};
		\node [style=none, label={[xshift=-2.5em]left:5x}] (33) at (0,0)[xshift=12em, yshift=-3.3em] {};
		\node [style=none] (34) at (0,0)[xshift=12em, yshift=-16.0em] {};		

		\node [style=container box] (35) at (0,0)[xshift=24em, yshift=0em] {};
		\node [style=box] (25) at (0,0)[xshift=24em, yshift=-1em, label={[yshift=-1em, align=center]\titleface{Downsampling}}] {};
		\node [style=threelines, label={[align=center, anchor=north]Conv2D\\\small{k{:} 3x3, s{:} 2x2,}\\\small{$C_{\text{out}}${:} $C_{\text{in}}$}}] (36) at (0,0)[xshift=24em, yshift=-4.5em] {};
		\node [style=oneline, fill=coolor4, fill opacity=1] (37) at (0,0)[xshift=24em, yshift=-10em] {Batch Norm.};
		\node [style=oneline, fill=coolor2, fill opacity=1] (38) at (0,0)[xshift=24em, yshift=-13em] {ReLU};
		\node [style=innercontainer, minimum height=11em] (78) at (0,0)[xshift=24em, yshift=-4.0em] {};
		\node [style=none, label={[xshift=-2.5em]left:2x}] (39) at (0,0)[xshift=24em, yshift=-3.3em] {};
		\node [style=none] (40) at (0,0)[xshift=24em, yshift=-16.0em] {};		

		\node [style=container box] (62) at (0,0)[xshift=36em, yshift=0em] {};	
		\node [style=box] (26) at (0,0)[xshift=36em, yshift=-1em, label={[yshift=-1em, align=center]\titleface{Channel Downscaling}}] {};
		\node [style=threelines, label={[align=center, anchor=north]Conv2D\\\small{k{:} 3x3, s{:} 1x1,}\\\small{$C_{\text{out}}${:} $\frac{1}{2}C_{\text{in}}, 3C$}}] (63) at (0,0)[xshift=36em, yshift=-4.5em] {};
		\node [style=oneline, fill=coolor4, fill opacity=1] (64) at (0,0)[xshift=36em, yshift=-9em] {Batch Norm.};
		\node [style=oneline, fill=coolor2, fill opacity=1] (65) at (0,0)[xshift=36em, yshift=-11em] {ReLU};
		\node [style=innercontainer, minimum height=9em] (79) at (0,0)[xshift=36em, yshift=-4.0em] {};
		\node [style=none, label={[xshift=-2.5em]left:2x}] (66) at (0,0)[xshift=36em, yshift=-3.3em] {};
		\node [style=none] (91) at (0,0)[xshift=36em, yshift=-15.5em] {};
		\node [style=none] (92) at (0,0)[xshift=41.5em, yshift=-11.25em] {};
		
		\node [style=oneline, fill=coolor5, fill opacity=1] (68) at (0,0)[xshift=36em, yshift=-13.5em] {Linear};
		\node [style=oneline, fill=coolor2, fill opacity=1] (69) at (0,0)[xshift=36em, yshift=-15.5em] {ReLU};
		\node [style=oneline, fill=coolor5, fill opacity=1] (70) at (0,0)[xshift=36em, yshift=-17.5em] {Linear};
		\node [style=none, label={[align=center]below:\Large{$\hat{\bm\eta}$}}] (72) at (0,0)[xshift=36em, yshift=-20em] {};

		\node [style=container box] (52) at (0,0)[xshift=48em, yshift=0em] {};
		\node [style=box] (28) at (0,0)[xshift=48em, yshift=-1em, label={[yshift=-1em, align=center]\titleface{Model Order}}] {};
		\node [style=threelines, label={[align=center, yshift=-4.0em]Conv2D\\\small{k{:} 3x3, s{:} 1x1,}\\\small{$C_{\text{out}}${:} 4}}] (82) at (0,0)[xshift=48em, yshift=-4.5em] {};
		\node [style=oneline, align=center, fill=coolor4, fill opacity=1] (83) at (0,0)[xshift=48em, yshift=-9em] {Batch Norm.};
		\node [style=oneline, align=center, fill=coolor2, fill opacity=1] (84) at (0,0)[xshift=48em, yshift=-11em] {ReLU};
		\node [style=none] (85) at (0,0)[xshift=48em, yshift=-3em] {};
		
		\node [style=oneline, align=center, fill=coolor5, fill opacity=1] (75) at (0,0)[xshift=48em, yshift=-13.5em] {Linear};
		\node [style=oneline, fill=coolor2, fill opacity=1] (76) at (0,0)[xshift=48em, yshift=-15.5em] {ReLU};
		\node [style=oneline, align=center, fill=coolor5, fill opacity=1] (77) at (0,0)[xshift=48em, yshift=-17.5em] {Linear};
		\node [style=none, label={[align=center]below:\Large{$\hat{\bm\rho}$}}] (55) at (0,0)[xshift=48em, yshift=-20em] {};

	\end{pgfonlayer}
	\begin{pgfonlayer}{edgelayer}
		\draw [style=thickarrow] (33) to (29);
		\draw [style=thickarrow] (66) to (63);
		\draw [style=thickarrow] (39) to (36);
		
		\draw [style=thickarrow] (29) to (30);
		\draw [style=thickarrow] (30) to (31);
		
		\draw [style=thickarrow] (36) to (37);
		\draw [style=thickarrow] (37) to (38);
		\draw [style=thickarrow] (31) to (34);
		\draw [style=thickarrow] (38) to (40);
		
		\draw [style=thick line] (63) to (64);
		\draw [style=thick line] (64) to (65);
		
		\draw [style=thick line] (68) to (69);
		\draw [style=thick line] (69) to (70);
		
		\draw [style=thick line] (75) to (76);
		\draw [style=thick line] (76) to (77);
		\draw [style=thickarrow] (77) to (55);
		\draw [style=thickarrow] (65) to (68);
		
		\draw [style=thick line] (82) to (83);
		\draw [style=thick line] (83) to (84);
		\draw [style=thickarrow] (83) to (75);
		
		\draw [style=thickarrow] (69) to (72);
		\draw [style=thickarrow] (85) to (82);
		
		\draw [style=thickarrow, ultra thick] (32.east) to (35.west);
		\draw [style=thickarrow, ultra thick] (35.east) to (62.west);
		
		\draw [thick, densely dashed, ultra thick] ([yshift=-1em]35.east) to ([yshift=-1em]62.west);
		\draw [style=thickarrow, densely dashed, ultra thick] ([yshift=-1em]62.east) to ([yshift=-1em]52.west);
	\end{pgfonlayer}
\end{tikzpicture}
    \end{scaletikzpicturetowidth}
    \caption{The architecture of our \gls{cnn} uses convolutional layers to perform upscaling, downsampling and downscaling. The encoded parameters in $\bm\eta$ and model order $\bm\rho$ (see \Cref{sec:neuralnetwork:offgrid} and \Cref{sec:neuralnetwork:architecture}, respectively) are estimated from dense layers to the downscaled or downsampled result.}
    \label{fig:architecture}
\end{figure*}
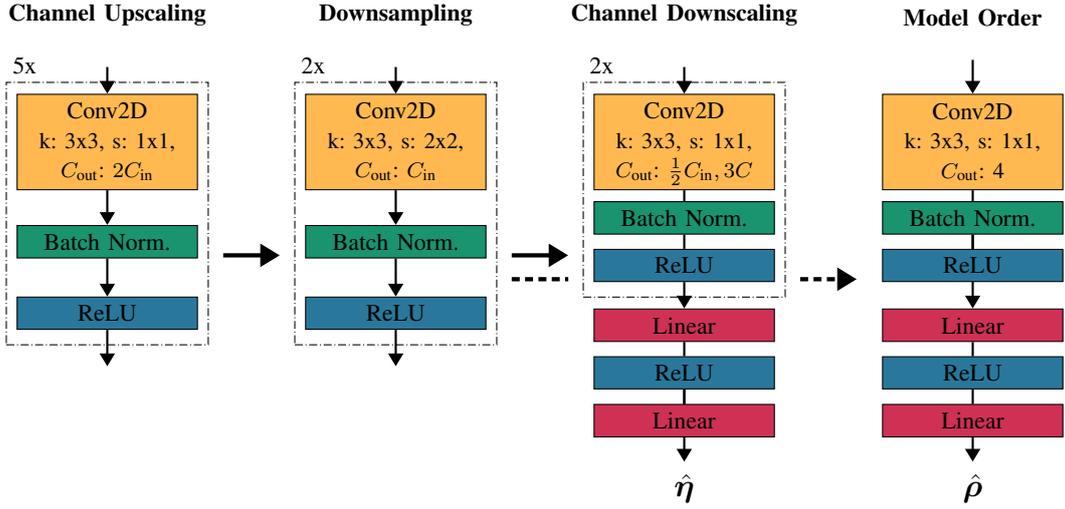

\section{Signal Model}\label{sec:signalmodel}
Our task is retrieving the spectral paths of deterministic paths from frequency and time samples of a radio channel, i.e., their propagation delays $\tau$ and Doppler-shifts $\alpha$.

We model the wireless channel transfer-function measurement of bandwidth $B$ with $N_f \in \mathbb{N}$ frequency samples and $N_t \in \mathbb{N}$ snapshots and employ the narrowband assumption since $B \ll f_c$.
We denote the sampled observation in complex baseband ($f_c = 0$) by $\bm S$.
The sampling process is characterized by the sampling intervals in frequency $\Delta f > 0$ and time $\Delta t > 0$ with $\bm S$ sampled $N_f, N_t \in \N$ times at
\begin{equation}
    f_k = f_0 + k \cdot \Delta f,
    t_l = t_0 + l \cdot \Delta t
\end{equation}
where $k = 0,...,N_f-1$, $l = 0,...,N_t-1$, $f_0 = - B/2$, and $t_0 = 0$.
Therefore, the discrete signal model $\bm S \in \C^{N_f \times N_t}$ is formulated as
\begin{align}
    S_{k,l}(\bm \gamma, \bm \tau, \bm \alpha) = \sum_{p=1}^{P} \gamma_p \exp{(-2j\pi f_k \tau_p)} \exp{(2j\pi t_l \alpha_p)} \label{eq:signal_discrete},
\end{align}
where the index $p = 1,...,P$ denotes the path index and $\bm\gamma \in \C^{P}$, $\bm\tau \in \R^{P}$, $\bm\alpha \in \R^{P}$ contain the corresponding complex weights, delays, and Doppler-shifts, respectively.
The noisy observation $\bm Y \in \C^{N_f \times N_t}$ is then formulated as
\begin{equation}\label{eq:observation_discrete}
    \bm Y = \bm S(\bm\gamma, \bm\tau, \bm\alpha) + \bm N ,
\end{equation}
where $\bm N \in \C^{N_f \times N_t}$ is a complex, zero-mean Gaussian noise process with variance $\sigma^2$.

From \eqref{eq:observation_discrete}, it follows that the task of estimating $P$, $\bm \tau$ and $\bm \alpha$ from $\bm Y$ constitutes a joint model order selection and harmonic retrieval problem.

\section{Neural Network}\label{sec:neurealnetwork}
The goal of the presented approach is to use a deep convolutional neural network to estimate $\bm\tau$ and $\bm\alpha$ from the sampled observations in $\bm Y$.
This section introduces the preprocessing applied to the data $\bm Y$, the off-grid parameter encoding used for the labels in the supervised training, and the network architecture.
\subsection{Preprocessing}\label{sec:neuralnetwork:preprocessing}
Our preprocessing stage aims to provide the neural network with informative and diverse input values via a two-step approach.\par
In the first step, multiple views of the data $\bm Y$ are created by filtering with $N_W$ windows and stacking the results into $\bm Y_W \in \C^{N_W \times N_f \times N_t}$. 
The motivation for the multi-window approach is to obtain different data realizations of the same data samples, i.e., retrieve different information from the same samples.
Rectangular windows achieve the maximum \gls{snr} and provide a narrow pulse shape but also result in high sidelobes after the \gls{dft}, potentially introducing ghost paths.
Other filters, such as Hann-windows, reduce the sidelobes, i.e., the probability of ghost paths, but increase the mainlobe width and usually resulting in higher estimation variances.
Hence, the choice for a window function is usually application specific and bears trade-offs.
However, a \gls{cnn} can process multiple views of the same data in parallel, similar to color channels of an image.
With this approach, diverse views can be exploited for more robust estimates.
We used $N_W=8$ different windowing functions, i.e., a Tukey, Taylor, Chebyshev, Blackman, Flat Top, Cosine, Hann, and the Rectangular window.\par
The second step is to apply a 2D-\gls{dft} over the last two dimensions, transforming it to the target parameter domain in delay- and Doppler, denoted as $\bm Y_1 \in \C^{N_W \times N_f \times N_t}$. 
To obtain real-valued numbers for the training, we employ four mapping functions
\begin{align*}
    f_1(\bm Y_1) = \Re(\bm Y_1),
    f_2(\bm Y_1) = \Im(\bm Y_1), \\
    f_3(\bm Y_1) = \log_{10}(\vert \bm Y_1 \vert),
    f_4(\bm Y_1) = \angle(\bm Y_1)
\end{align*}
to map the complex values in $\bm Y_1$ to the real-valued \gls{cnn} input data $\bm Y_2 \in \R^{4 \cdot N_W \times N_f \times N_t}$. 
Here, $\Re$ and $\Im$ denote the real and imaginary parts, respectively, and $\vert \cdot \vert$ and $\angle$ denote the absolute value and phase of a complex number. 
Even though $f_1$ and $f_2$ contain the same information as $f_3$ and $f_4$, our experiments showed, the simultaneous usage adds useful diversity for the \gls{cnn}.
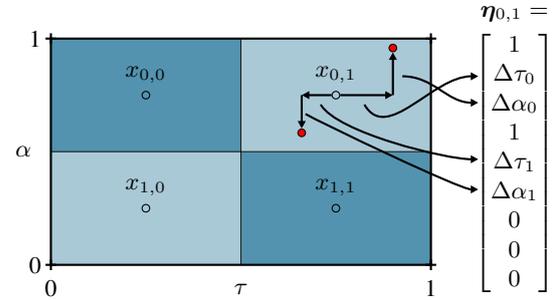
\begin{figure}[t]
    \centering

\tikzstyle{grid}=[fill=none, draw=black, shape=rectangle, minimum width=5cm, minimum height=3cm, thick]
\tikzstyle{xtick}=[fill=white, draw=black, shape=rectangle, minimum height=4pt, minimum width=0pt, inner sep=0pt, thick]
\tikzstyle{ytick}=[fill=white, draw=black, shape=rectangle, minimum height=0pt, minimum width=4pt, inner sep=0pt, thick]
\tikzstyle{cell}=[fill=none, draw=black, shape=rectangle, minimum width=2.5cm, minimum height=1.5cm]
\tikzstyle{param}=[fill=red, draw=black, shape=circle, minimum width=0.1cm, minimum height=0.1cm, inner sep=0pt]
\tikzstyle{center}=[fill=none, draw=black, shape=circle, minimum width=0.1cm, minimum height=0.1cm, inner sep=0pt]
\tikzstyle{arrow}=[->, >={Triangle[length=1mm, width=1mm]}, thick]

\begin{tikzpicture}
	\begin{pgfonlayer}{nodelayer}
		\node [style=cell, fill=coolor2, opacity=0.8] (11) at (-1.25, 0.75) {};
		\node [style=cell, fill=coolor2, opacity=0.4] (12) at (1.25, 0.75) {};
		\node [style=cell, fill=coolor2, opacity=0.8] (13) at (1.25, -0.75) {};
		\node [style=cell, fill=coolor2, opacity=0.4] (14) at (-1.25, -0.75) {};	
		\node [style=grid] (0) at (0, 0) {};
		\node [style=xtick, label={left:1}] (2) at (-2.5, 1.5) {};
		\node [style=xtick, label={below:1}] (3) at (2.5, -1.5) {};
		\node [style=xtick, label={below:0}, label={left:0}] (4) at (-2.5, -1.5) {};
		\node [style=xtick] (5) at (2.5, 1.5) {};
		\node [style=ytick] (7) at (2.5, 1.5) {};
		\node [style=ytick] (8) at (-2.5, 1.5) {};
		\node [style=ytick] (9) at (-2.5, -1.5) {};
		\node [style=ytick] (10) at (2.5, -1.5) {};
		\node [style=param] (15) at (0.8, 0.25) {};
		\node [style=none, below] (16) at (0, -1.75) {$\tau$};
		\node [style=none, left] (17) at (-2.75, 0) {$\alpha$};
		\node [style=param] (18) at (2, 1.375) {};
		\node [style=center, label={above:$x_{0,1}$}] (20) at (1.25, 0.75) {};
		\node [style=center, label={above:$x_{0,0}$}] (21) at (-1.25, 0.75) {};
		\node [style=center, label={above:$x_{1,0}$}] (22) at (-1.25, -0.75) {};
		\node [style=center, label={above:$x_{1,1}$}] (23) at (1.25, -0.75) {};
		\node [style=none] (24) at (2, 0.75) {};
		\node [style=none] (26) at (0.8, 0.75) {};
		\node [style=none] (30) at (1.25, 1.5) {};
		\node [style=none] (31) at (0.75, 1) {};
		\node [style=none, label={[]$\bm \eta_{0,1}=$}] (32) at (3.6, -0.15) {$\begin{bmatrix} 1 \\ \Delta\tau_0 \\ \Delta\alpha_0 \\ 1 \\ \Delta\tau_1 \\ \Delta\alpha_1 \\ 0 \\ 0 \\ 0\end{bmatrix}$};
		\node [style=none] (28) at (2.125, 1) {}; %
		\node [style=none] (29) at (1.625, 0.625) {};
		\node [style=none] (34) at (3.125, 0.65) {};
		\node [style=none] (36) at (3.125, 1) {};
		\node [style=none] (37) at (4.75, 0.75) {}; 
		\node [style=none] (38) at (0.85, 0.5) {};
		\node [style=none] (39) at (3.125, -0.5) {};
		\node [style=none] (40) at (1.05, 0.625) {};
		\node [style=none] (41) at (3.125, -0.1) {};
		
		\draw [style=arrow] (20) to (24.center);
		\draw [style=arrow] (24.center) to (18);
		\draw [style=arrow] (20) to (26.center);
		\draw [style=arrow] (26.center) to (15);
		\draw [style=arrow, in=180, out=0] (28.center) to (34.center);
		\draw [style=arrow, in=-180, out=-60] (29.center) to (36.center);
		\draw [style=arrow, in=180, out=-30, looseness=0.25] (38.center) to (39.center);
		\draw [style=arrow, in=180, out=-60, looseness=0.50] (40.center) to (41.center);		
	\end{pgfonlayer}
\end{tikzpicture}
    \caption{Example for the label encoding with $C=3$. The path parameters (\protect\reddot) are encoded relative the closest cell-centroid  $\bm\eta_{i,j}$.}
    \label{fig:labels}
\end{figure}
\subsection{Off-Grid Parameter Encoding}\label{sec:neuralnetwork:offgrid}
The labels for the supervised learning task are provided by a grid-relative parameter encoding. 
To this end, we divide the delay-Doppler domain into a few grid-cells. 
In each grid-cell, we aim to estimate the number of paths in that cell and their respective deviations from the cell center.
Hence, the parameter values are encoded relative to the grid cell centers.
Our encoding is a task-specific modification of \cite{redmon_you_2016}.
It consists of three steps: parameter normalization, cell assignment, and relative parameter encoding.\par
Let $C$ be the maximum number of paths in a single cell.
The normalization maps the parameters into a range between 0 and 1, such that $\bm\tau$ and $\bm\alpha$ are in the interval of $\tau_p \subset \left[0, 1\right)$ and $\alpha_p \subset \left[0, 1\right)$.
Next, we define a set of $I \cdot J$ cell centers $\bm x \in (0,1)^{I \times J}$, which each define a non-overlapping rectangular covering of $[0,1) \times [0,1)$.
Paths are mapped to the cells based on the shortest $\ell_2$-distance to the cell centroids $x_{i,j}$.
To encode the paths' positions and number in each cell, we define vectors $\bm \eta_{i,j} \in \R^{3 \cdot L_{\rm max}}$ as
\begin{equation}
    \bm\eta_{i,j} = \left[
    \mu_{1}^{[i,j]}, \Delta\tau_{1}^{[i,j]}, \Delta\alpha_{1}^{[i,j]}, \hdots, \mu_{C}^{[i,j]}, \Delta\tau_{C}^{[i,j]}, \Delta\alpha_{C}^{[i,j]}
    \right]^T
\end{equation}
$\Delta\tau_c^{[i,j]} = \Norm{\tau_{c} - x_{i,j}}_2$ and $\Delta\alpha_c = \Norm{\alpha_{c} - x_{i,j}}_2$ denote the Euclidean distance between the cell centroid $\eta_{i,j}$ and the respective parameters.
In line with best practices for training, $\Delta\tau_c$ and $\Delta\alpha_c$ are normalized by the cell width and shifted based on the cells' centers, such that $\Delta\tau_c, \Delta\alpha_c \subset \left[0, 1\right)$.
Note, that the model order $P$ can be expressed as $P = \sum_{i,j,c}^{I,J,C} \mu_{c}^{[i,j]}$ and the maximum number of encodable paths is $C\cdot I \cdot J$.

As the number of paths in each cell can vary, $\mu_{c}^{[i,j]} = \{0, 1\}$ indicates if the path with encoding $\Delta\tau_{c}^{[i,j]}$ and $\Delta\alpha_{c}^{[i,j]}$ is an estimate ($\mu_{c}^{[i,j]} = 1$) or empty ($\mu_{c}^{[i,j]} = 0$).
This enables computing a coupled loss during training, as detailed in \Cref{sec:neuralnetwork:lossfunctions}, for arbitrary cell assignments.
To enforce a predictable ordering of paths in each cell, the paths are sorted from $c=1...C$ with descending magnitude $\gamma_p$, and unassigned parameters are labeled as \num{0} (see \Cref{fig:labels}).
The result of the encoding is a 3D array $\bm\eta \in \R^{I \times J \times 3 \cdot C}$, which structures the desired prediction results of our \gls{cnn}.
\subsection{Network Architecture}\label{sec:neuralnetwork:architecture}
Our network architecture is split into four stages represented in \Cref{fig:architecture}.
The first stage passes the input through \num{5} blocks of 2D convolutional layers.
Each block consists of a 2D convolutional layer, followed by Batch-Normalization and a \gls{relu} activation function.
The convolutional layers are parameterized to preserve the data shape but double the number of channels after each block.\par
The second stage performs downsampling via convolutional layers with a stride of \num{2}, which reduces the data dimension by \num{2} with each block.
In this stage, the number of channels is preserved.\par
Stage three achieves the parameter predictions.
First, the number of channels is reduced by two blocks of convolutional layers, followed by two \gls{fc} layers interleaved by a \gls{relu} activation function.
The relative parameter estimates $\bm\eta$ (from \Cref{sec:neuralnetwork:offgrid}) are encoded in the output of the final \gls{fc} layer.\par
The fourth stage is used to predict the number of paths $P$ (model order) based on the results of the second stage.
It uses a single convolutional block followed by two \gls{fc} layers interleaved by a \gls{relu} activation function.
Its output is $\hat{\bm\rho}$, a one-hot encoded vector of the model order estimate $\hat{P}$.
\begin{figure*}[t]
    \input{figures/inference_example.tex}
\end{figure*}
\subsection{Loss Functions and Training}\label{sec:neuralnetwork:lossfunctions}
Our approach uses multiple loss functions combined in a weighted sum.
The first summand is the loss for the model order estimate $\hat{\bm\rho}$.
It uses the well-known \gls{bce} loss for the one-hot encoded values.
\begin{align}
    \mathfrak{L}_{0} = \hat{\rho} \cdot \log(\rho) + (1-\hat{\rho}) \cdot \log(1-\rho) \label{eq:modelorderloss}
\end{align}
\begin{table}[b!]
    \centering
    \caption{Dataset summary and training hyperparameters.}
    \label{tab:settings}
    \begin{tabularx}{\linewidth}{Xp{4cm}}
        \toprule
        Name                                 & Value                                                                                  \\ \midrule
        \textbf{Datasets} & \mbox{}         \\
        Distribution $\tau_p$, $\alpha_p$    & $\mathfrak{U}_{[0,1]}$                                                                 \\

        Min. separation $\tau_p$, $\alpha_p$ & \num[exponent-mode=engineering, drop-zero-decimal=true]{0.003125}                      \\

        Magnitudes                           & $\mathfrak{U}_{[0.001, 1]}$                                                            \\
        Phases                               & $\mathfrak{U}_{[0,2\pi]}$                                                              \\
        SNR                                  & \qtyrange{0}{50}{\decibel}                                                             \\
        Number of Paths                      & $\mathfrak{U}_{[1,20]}$                                                                \\
        Trainingset Size                     & \num[exponent-mode=engineering, drop-zero-decimal=true]{400000}                        \\
        Validationset Size                   & \num{1000}                                                                             \\
        Testset Size                         & \num{4000}                                                                             \\ \midrule
        \textbf{Training}                    &                                                                                        \\
        Optimizer                            & Adam \cite{kingma_adam_2014}, $\gamma=0.0003$, \newline $\beta_1=0.9$, $\beta_2=0.999$ \\
        Mini-Batchsize                       & \num{32}                                                                               \\
        Epochs                               & 20                                                                                     \\
        Trainable Parameters                 & \num[exponent-mode=engineering, drop-zero-decimal=true]{25e6} for $N_f=N_t=64$         \\\bottomrule
    \end{tabularx}
\end{table}
The loss for the parameter estimates $\bm\eta$ utilizes a masked \gls{mse} loss function
\begin{align}
    \mathfrak{L}_{1} = \sum_{i,j=1}^{I,J}\sum_{c=1}^{C} \left(
    \sigma \left(\hat{\mu}_c^{[i,j]}\right)
    \cdot
    \Norm{\begin{bmatrix}
                      \Delta\hat{\tau}_{c}^{[i,j]} \\
                      \Delta\hat{\alpha}_{c}^{[i,j]}
                  \end{bmatrix}
            -
            \begin{bmatrix}
                \Delta\tau_{c}^{[i,j]} \\
                \Delta\alpha_{c}^{[i,j]}
            \end{bmatrix}}_1
    \right)^2 \label{eq:parameterloss},
\end{align}
where $\sigma(\cdot)$ represents the sigmoid function, and $\Hat{\cdot}$ marks the predictions.
As mentioned earlier, $\mu_{c}^{[i,j]}$ is used to weight the parameter estimates $\Delta\tau_{c}^{[i,j]}$ and $\Delta\alpha_{c}^{[i,j]}$ during loss calculation.
Its effect becomes apparent by inspecting the limits of the $\sigma(\cdot)$ function, as
\begin{enumerate}
    \item $\lim_{x \to \infty}\sigma(x) = 1$, causing the \gls{mse} of the corresponding predictions to contribute to $\mathfrak{L}_{1}$.
    \item $\lim_{x \to -\infty}\sigma(x) = 0$, causing the \gls{mse} of the corresponding predictions to \textbf{not} contribute to $\mathfrak{L}_{1}$.
\end{enumerate}
Hence, predicting a negative value for $\mu_c^{[i,j]}$ causes $\sigma(\mu_c^{[i,j]})\Delta\hat{\tau_{c}}^{[i,j]} \approx 0$ and $\sigma(\mu_c^{[i,j]})\Delta\hat{\alpha_{c}}^{[i,j]} \approx 0$ and hence close to the corresponding \num{0} in the labels.\par

Finally, both loss components are combined in a weighted sum via
\begin{align}
    \mathfrak{L_{\text{total}}} = \mathfrak{L_{0}} + \beta \cdot \mathfrak{L}_{1}.\label{eq:totalloss}
\end{align}
For our experiments, we manually selected $\beta = 4$ to ensure both losses equally contribute to the learning.
We note, that $\beta$ is likely not optimal and choosing it constitutes a multi-objective optimization problem for further study.

\subsection{Training}\label{sec:neuralnetwork:training}
The three synthetic datasets, a training, validation, and test set, were created by sampling random values for the signal parameters.
\Cref{tab:settings} contains a comprehensive summary of the respective settings for the dataset and training hyperparameters.\par
Each sample in the dataset contains a random number of \numrange{1}{20} specular paths.
The complex path amplitudes $\gamma$ contain random phases, and their magnitudes are uniformly spread across the range of \qtyrange{0}{-30}{\decibel}.
To prevent overfitting, the measurement noise $\bm N$ is generated randomly for every snapshot with a random noise variance $\sigma$, such that the \gls{snr} is in the range of \qtyrange{0}{50}{\decibel}.
We use a uniform distribution in the linear domain, such that \qty{90}{\percent} of samples have a \gls{snr} \qty{< 10}{\decibel}.

\section{Analysis}\label{sec:analysis}
In order to get a complete estimate of the parameters in \eqref{eq:observation_discrete}, we retrieved estimates $\hat{P}$, $\hat{\bm \tau}$ and $\hat{\bm\alpha}$ and based on these used the \gls{blue}, i.e., least-squares, to attain an estimate for the linear weights $\hat{\bm \gamma}$.

To illustrate the results obtained from our \gls{cnn}, \Cref{fig:example} shows a single, hand-picked sample from the validationset passed through the network at different \glspl{snr}. 
Not only do we analyze the raw parameter estimates of the network, but we also use these estimates to warm-start a gradient iteration
\begin{equation}\label{gradient_steps}
    (\bm \gamma^{k+1}, \bm \tau^{k+1}, \bm \alpha^{k+1}) =
    (\bm \gamma^{k}, \bm \tau^{k}, \bm \alpha^{k}) - \varepsilon^k \bm z^k
\end{equation}
with descent direction
\[
    \bm z^k = \left[\left(\bm F^k\right)^{-1} \cdot \bm J^k\right](\bm \gamma^{k}, \bm \tau^{k}, \bm \alpha^{k}),
\]
which essentially defines a second-order Gauss-Newton scheme, since $\bm F$ is the Fisher-Information matrix and $\bm J$ is the Jacobian matrix of the negative $\log$-likelihood function based on the assumption that $\bm N$ is a Gaussian random variable, which reads as
\begin{equation}\label{llf}
    \lambda(\bm \gamma, \bm \tau, \bm \alpha) = \frac{1}{\sigma^2}\Norm{\bm Y - \bm S(\bm\gamma, \bm\tau, \bm\alpha)}_F^2.
\end{equation}

To provide an assessment of the estimation performance of our approach, we compare it to a periodogram-based peak search, i.e., \gls{dft}, which is inherently grid-limited, and the high-resolution \gls{rimax}~\cite{richter_estimation_2005} based on \gls{ml}.
The model order required for the peak search is obtained from the \gls{edc}~\cite{zhao_detection_1986}.
We provide a comparison in terms of the \gls{mse} for the estimated parameters and the respective model order error.

\begin{figure}[t]
    \begin{center}
        \input{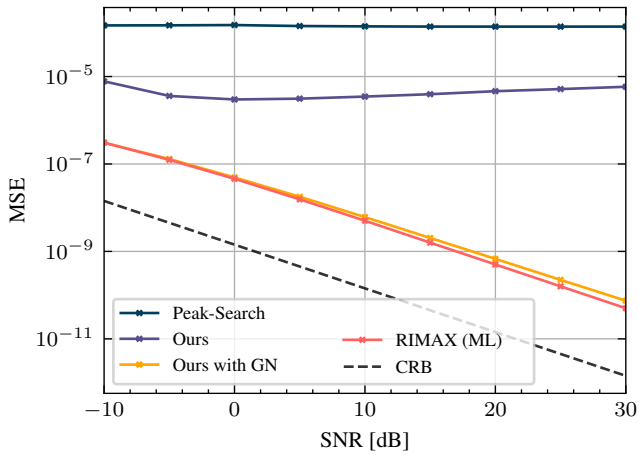}
    \end{center}
    \caption{Comparison of the \gls{mse}. Our method can outperform the periodogram method in terms of accuracy but is surpassed by the high accuracy of the \gls{rimax} algorithm. With \num{10} additional gradient-steps on the likelihood function using the estimates from our approach as initialization, we can achieve similar performance to the \gls{ml} method at significantly lower computation times.}
    \label{fig:mse}
\end{figure}

\Cref{fig:mse} shows that our approach can overcome the grid limitation and outperforms the periodogram-based method.
However, the raw estimates' accuracy of our approach also saturates, albeit at a lower \gls{mse}.
As expected, the high-resolution estimator's results align with the predictions of the \gls{crb} with increasing \gls{snr}.\footnote{The \gls{mse} is computed only for those estimated parameters with a match in the groundtruth within a 1/N distance. We confirmed, that \gls{rimax} reaches the \gls{crb} for a single path scenario.} 
When using the estimates provided by our approach to initialize \eqref{gradient_steps}, the estimation accuracy for higher \glspl{snr} improves significantly from only a \num{10} iterations of \eqref{gradient_steps}. 
This is a highly promising feature of our architecture since the \emph{joint} estimates are close enough to the global minimum of \eqref{llf} such that \eqref{gradient_steps} is very likely to converge to the true solution.
This starkly contrasts algorithms like \gls{rimax}, which only add single sources one by one to the set of estimates and carry out iterative refinement between two newly added sources, known as successive interference cancellation.
In our case, we can initialize the iteration much more efficiently with a single forward from the network but still accurate enough for the gradient iteration to converge.\par
Apart from the accuracy, the computational complexity of the algorithms is also of interest.
As an assessment of the computational complexity of the \gls{rimax} estimator is not straightforward due to the use of iterative numerical methods, we assess the runtime per sample on an identical system.
The periodogram approach ranks fastest with an average of \SI{3}{\milli\second}, followed by our approach with \SI{19}{\milli\second}.
When combining our method with $10$ iterations of \eqref{gradient_steps} we average at \SI{60}{\milli\second}, while the \gls{ml} algorithm \gls{rimax} requires around \SI{11.9}{\second} on average.
It highlights that our approach addresses applications requiring fast fixed-clock estimates, where the accuracy-runtime trade-off can be regulated by the number of gradient iterations.

\begin{figure}[t]
    \begin{center}
        \input{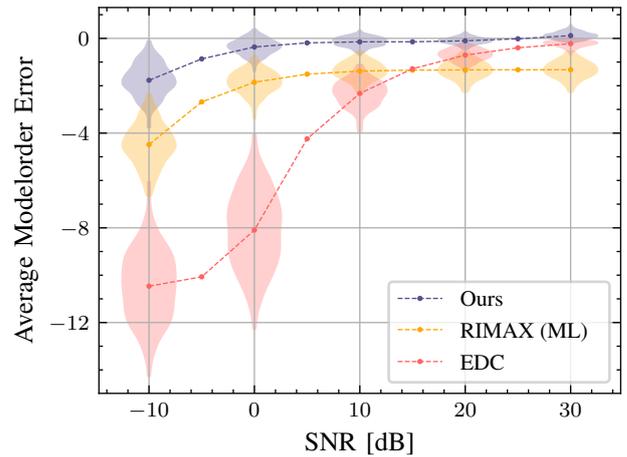}
    \end{center}
    \caption{Average difference of the true and estimated model order for the simulations in \Cref{fig:mse}. Our method outperforms both \gls{edc} and the maximum-likelihood approach.}
    \label{fig:modelordererror}
\end{figure}

Regarding the model order estimation, we compare our approach to \gls{edc} and the model order extracted from the \gls{ml} estimates.
The results are illustrated in \Cref{fig:modelordererror} and reaffirm the findings of previous publications \cite{izacard_data-driven_2019, barthelme_machine_2021}, where it is shown that neural networks can reliably predict the number of sources in a signal.
Our approach consistently achieves the best results across the studied \gls{snr} range.
\Gls{edc} and \gls{rimax} achieves similar performance at \glspl{snr} \SI{>20}{\decibel}, but underestimates the model order for smaller \gls{snr}.
Overall, this result highlights the model order estimation capabilities of our approach, particularly in the challenging low-\gls{snr} domain.

\section{Conclusion}\label{sec:conclusion}
Our work introduces a new approach for combined, two-dimensional harmonic retrieval and model order estimation using a \gls{cnn}.
Compared to recent approaches in the field, it uses a cell-based representation of spectral parameters for the prediction and, therefore, can estimate parameters directly via regression instead of classification.
Regarding estimation accuracy, it outperforms on-grid periodogram-based approaches.
Most interestingly, the estimates of up to \num{20} paths can be used to warm-start a second-order gradient iteration of the highly non-convex likelihood function. 
With some further refinement, the architecture can initialize a well-performing \gls{ml}, hence approximately delivering all the advantageous properties, like statistical consistency and efficiency. 
Especially compared to \cite{barthelme_machine_2021}, this is an improvement in terms of parameter dimension, quantity, and data complexity.

Additionally and in line with previous work, it demonstrates superior model order estimation, especially in the low-\gls{snr} regime.
Our approach is well-suited for time-constraint harmonic retrieval tasks because of its relatively low runtime compared to high-resolution methods.

Due to the \gls{cnn} structure of our approach, it scales well to more than two dimensions. 
Ultimately, this should allow the \gls{cnn} processing of \gls{mimo} measurements and hence full channel sounding data, including spatial measurements, similar to \cite{richter_estimation_2005}.
Then, the preprocessing must be extended with realistic antenna beampatterns by a suitable beamspace transformation such as the \gls{eadf}, see \cite{landmann2004EADF}.
Moreover, an ablation study should quantify the performance impacts of the individual architecture blocks.
Similarly, processing real measurement data will help us understand to what degree the approach is affected by measurement system imperfections and model mismatch.
Further opportunities are the applicability to wideband channel data, where the narrowband assumption is no longer satisfied, leading to dispersion in delay and Doppler domains and coupling of the two parameters.
%

\section*{Acknowledgment}
The authors acknowledge the financial support by the Federal Ministry of Education and Research of Germany in the project “Open6GHub” (grant number: 16KISK015), “KOMSENS-6G” (grant number: 16KISK125), and DFG project HoPaDyn with Grant-No. TH 494/30-1.

\printbibliography

@inproceedings{izacard_data-driven_2019,
      title={Data-Driven Estimation of Sinusoid Frequencies}, 
      author={Izacard, Gautier and Mohan, Sreyas and Fernandez-Granda, Carlos},
      booktitle = {Advances in Neural Information Processing Systems},
      year = {2019},
      pages = {5127--5137},
      volume = {32},
      url = {https://proceedings.neurips.cc/paper/2019/file/d0010a6f34908640a4a6da2389772a78-Paper.pdf},
}

@article{zhao_detection_1986,
	title = {On detection of the number of signals in presence of white noise},
	volume = {20},
	issn = {0047259X},
	url = {https://linkinghub.elsevier.com/retrieve/pii/0047259X86900175},
	doi = {10.1016/0047-259X(86)90017-5},
	language = {en},
	number = {1},
	journal = {Journal of Multivariate Analysis},
	author = {Zhao, L.C. and Krishnaiah, P.R. and Bai, Z.D.},
	month = oct,
	year = {1986},
	pages = {1--25},
}

@article{barthelme_machine_2021,
	title = {A {Machine} {Learning} {Approach} to {DoA} {Estimation} and {Model} {Order} {Selection} for {Antenna} {Arrays} {With} {Subarray} {Sampling}},
	volume = {69},
	issn = {1053-587X, 1941-0476},
	url = {https://ieeexplore.ieee.org/document/9432742/},
	doi = {10.1109/TSP.2021.3081047},
	journal = {IEEE Transactions on Signal Processing},
	author = {Barthelme, Andreas and Utschick, Wolfgang},
	year = {2021},
	pages = {3075--3087},
}

@article{naseri_machine_2022,
	title = {Machine {Learning}-{Based} {Angle} of {Arrival} {Estimation} for {Ultra}-{Wide} {Band} {Radios}},
	issn = {1558-2558},
	doi = {10.1109/LCOMM.2022.3167020},
	journal = {IEEE Communications Letters},
	author = {Naseri, Mostafa and Shahid, Adnan and Gordebeke, Gert-Jan and Lemey, Sam and Boes, Michiel and Van de Velde, Samuel and De Poorter, Eli},
	year = {2022},
	note = {Conference Name: IEEE Communications Letters},
	pages = {1--1},
}

@article{chen_robust_2022,
	title = {Robust {DoA} {Estimation} {Using} {Denoising} {Autoencoder} and {Deep} {Neural} {Networks}},
	issn = {2169-3536},
	doi = {10.1109/ACCESS.2022.3164897},
	journal = {IEEE Access},
	author = {Chen, Dawei and Shi, Shuo and Gu, Xuemai and Shim, Byonghyo},
	year = {2022},
	note = {Conference Name: IEEE Access},
	pages = {1--1},
}

@article{liu_super_2020,
	title = {Super resolution {DOA} estimation based on deep neural network},
	volume = {10},
	issn = {2045-2322},
	url = {http://www.nature.com/articles/s41598-020-76608-y},
	doi = {10.1038/s41598-020-76608-y},
	language = {en},
	number = {1},
	journal = {Scientific Reports},
	author = {Liu, Wanli},
	month = dec,
	year = {2020},
	pages = {19859},
}

@article{papageorgiou_deep_2020,
  author={Papageorgiou, Georgios K. and Sellathurai, Mathini and Eldar, Yonina C.},
  journal={IEEE Transactions on Signal Processing}, 
  title={Deep Networks for Direction-of-Arrival Estimation in Low SNR}, 
  year={2021},
  volume={69},
  number={},
  pages={3714-3729},
  doi={10.1109/TSP.2021.3089927}
}

@article{kingma_adam_2014,
  title={Adam: A Method for Stochastic Optimization},
  author={Diederik P. Kingma and Jimmy Ba},
  journal={CoRR},
  year={2015},
  volume={abs/1412.6980}
}

@inproceedings{redmon_you_2016,
	address = {Las Vegas, NV, USA},
	title = {You {Only} {Look} {Once}: {Unified}, {Real}-{Time} {Object} {Detection}},
	isbn = {978-1-4673-8851-1},
	shorttitle = {You {Only} {Look} {Once}},
	url = {http://ieeexplore.ieee.org/document/7780460/},
	doi = {10.1109/CVPR.2016.91},
	booktitle = {2016 {IEEE} {Conference} on {Computer} {Vision} and {Pattern} {Recognition} ({CVPR})},
	publisher = {IEEE},
	author = {Redmon, Joseph and Divvala, Santosh and Girshick, Ross and Farhadi, Ali},
	month = jun,
	year = {2016},
	pages = {779--788},
}

@phdthesis{richter_estimation_2005,
	address = {Ilmenau},
	type = {Thesis},
	title = {Estimation of {Radio} {Channel} {Parameters}: {Models} and {Algorithms}},
	url = {https://www.db-thueringen.de/servlets/MCRFileNodeServlet/dbt_derivate_00007407/ilm1-2005000111.pdf},
	language = {en},
	school = {Technische Universität Ilmenau},
	author = {Richter, Andreas},
	year = {2005},
}

@article{malioutov2005ssr_source_loc,
 author = {Malioutov, D. and Cetin, M. and Willsky, A.S.},
 journal = {IEEE Trans. Signal Process.},
 title = {A sparse signal reconstruction perspective for source localization with sensor arrays},
 date = {2005},
 volume = {53},
 number = {8},
 pages = {3010--3022},
 doi = {10.1109/tsp.2005.850882},
 source = {Crossref},
 url = {https://doi.org/10.1109/tsp.2005.850882},
 publisher = {Institute of Electrical and Electronics Engineers (IEEE)},
 issn = {1053-587X},
 year = {2005},
 month = aug,
}

@article{schmidt1986MUSIC,
  author    = {Schmidt, R.},
  journal   = {IEEE Trans. Antennas Propag.},
  title     = {Multiple emitter location and signal parameter estimation},
  date      = {1986},
  volume    = {34},
  number    = {3},
  pages     = {276--280},
  month     = mar,
  doi       = {10.1109/tap.1986.1143830},
  source    = {Crossref},
  url       = {https://doi.org/10.1109/tap.1986.1143830},
  publisher = {Institute of Electrical and Electronics Engineers (IEEE)},
  issn      = {0096-1973},
  year      = {1986}
}

@inbook{thomae2005multidim_cs,
  title     = {Multidimensional High-Resolution Channel Sounding Measurement},
  author    = {Thom{\"a}, {Reiner S.} and Markus Landmann and Andreas Richter and Uwe Trautwein},
  year      = {2005},
  language  = {English},
  isbn      = {977-5945-09-7},
  pages     = {241--270},
  editor    = {Thomas Kaiser and Andr{\'e} Bourdoux and Holger Boche and {Rodr{\'i}guez Fonollosa}, Javier and {Andersen Bach}, Jorgen and Wolfgang Utschick},
  booktitle = {Smart Antennas State of the Arts},
  publisher = {Hindawi Publishing Corporation},
  address   = {United States}
}

@inproceedings{landmann2004EADF,
  author    = {Landmann, M. and Galdo, G. Del},
  booktitle = {7th European Conference on Wireless Technology, 2004.},
  title     = {Efficient antenna description for {MIMO} channel modelling and estimation},
  date      = {2004},
  volume    = {},
  number    = {},
  pages     = {217--220},
  month     = oct
}

@article{roy1989esprit,
 author = {Roy, R. and Kailath, T.},
 journal = {IEEE Trans. Acoust. Speech Signal Process.},
 title = {{ESPRIT}-Estimation of signal parameters via rotational invariance techniques},
 date = {1989},
 volume = {37},
 number = {7},
 pages = {984--995},
 doi = {10.1109/29.32276},
 issn = {0096-3518},
 month = jul,
 source = {Crossref},
 url = {https://doi.org/10.1109/29.32276},
 publisher = {Institute of Electrical and Electronics Engineers (IEEE)},
 year = {1989},
}

@misc{cvx,
 author = {Grant, Michael and Boyd, Stephen},
 title = {Software},
 howpublished = {http://cvxr.com/cvx},
 month = apr,
 date = {2014},
 doi = {10.1002/9781119818243.app4},
 source = {Crossref},
 url = {https://doi.org/10.1002/9781119818243.app4},
 publisher = {Wiley},
 pages = {249--252},
 year = {2021},
}

@book{C,
 author = {Kernighan, Brian W. and Ritchie, Dennis M.},
 title = {The C Programming Language},
 date = {1988},
 isbn = {0131103709},
 publisher = {Prentice Hall Professional Technical Reference},
 edition = {2nd},
}

\end{document}